\documentclass[twocolumn,apjl]{openjournal}
\usepackage[utf8]{inputenc}
\usepackage{graphicx}
\usepackage{caption}
\usepackage{subcaption}
\usepackage{amsmath}
\usepackage{amssymb}
\usepackage{lineno}
\usepackage{bm}
\usepackage{latexsym}
\usepackage{url}
\usepackage{hyperref}

\begin{document}

\title{\vspace{-0.5cm} GCR Spectra Reconstructed with Neutron Monitor Yield Function and Artificial Neural Networks: Comparison of Two Methods\vspace{-1cm}}
\author{Stepan~A. Siruk$^{1*}$}
\author{Vladislav~V. Alekseev$^{2}$}
\author{Victor~A. Kuzminov$^{3}$}
\author{Rustam~F. Yulbarisov$^{1}$}
\author{Andrey~G. Mayorov$^{1}$}
\affiliation{$^{1}$Department of Experimental Nuclear Physics and Cosmophysics, National Research Nuclear University MEPhI, Moscow 115409, Russia}
\affiliation{$^{2}$Centre of Integrable Systems, P.G.~Demidov Yaroslavl State University, Yaroslavl 150003, Russia}
\affiliation{$^{3}$Department of Artificial Intelligence and Data Analysis, State Scientific Research Institute of Aviation Systems, Moscow 125319, Russia}

\thanks{$^*$E-mail: \href{mailto:SASiruk@mephi.ru}{SASiruk@mephi.ru}}


\begin{abstract}
We present a framework that reconstructs time-resolved galactic cosmic-ray (GCR) proton and helium energy spectra from the global neutron monitor network, providing data about GCR flux without direct satellite observations. Two methods are utilized and compared: a calibrated yield function plus force-field scheme and artificial neural networks trained on multi-station neutron monitor count rates coupled with heliophysical indices. The reconstructed spectral time series reproduce both large-scale solar-cycle modulation and short-term disturbances and extend to periods lacking daily spacecraft data, including 2006--2011 (consistent with PAMELA) and 2019--2022 (consistent with AMS-02 Bartels rotation averages). Artificial neural networks deliver excellent performance across energies, with markedly lower mean absolute percentage error and $\chi^2/\mathrm{dof}$ near unity. A thorough validation confirms robustness and establishes neutron monitors as an effective real-time GCR spectrometer that can be utilized for various purposes.
\end{abstract}



\section{Introduction \label{intro}}
Galactic cosmic rays (GCRs) are high-energy particles propagating through outer space. While passing through the heliosphere~--- a region of space where the solar wind plasma and magnetic field dominate over interstellar matter~--- GCRs interact with the medium. As a result, their differential energy spectra observed near the Earth differ significantly from the spectra outside the heliosphere, known as the local interstellar spectra (LIS). The intensity and nature of this interaction are determined by the current configuration of the heliosphere; therefore, the GCR flux varies over time. The dependence of GCR flux on solar activity is referred to as solar modulation~\citep{Potgieter_2013}. The most prominent manifestation of solar modulation is the long-term variation in the GCR flux, which occurs in antiphase with the 11-year solar activity cycle. In addition, periodic fluctuations are observed on shorter time scales, such as the 27-day GCR variations~\citep{Forbush_1938}, which arise due to the interaction of GCRs with structures known as corotating interaction regions~\citep{Heber_1999, Richardson_2018}. Another phenomenon associated with solar activity is the Forbush decrease (FD)~--- a rapid reduction in the GCR flux caused by the interaction with the fronts of coronal mass ejections propagating through interplanetary space~\citep{Forbush_1937}.

The study of GCR spectra and their variations is important from both fundamental and applied perspectives. GCRs carry valuable information about various processes occurring in the Sun, the interplanetary medium, and the Earth's magnetosphere~\citep{Abunina_2020-1, Danilova_2025}. Variations in the fluxes of secondary particles (produced by the interaction of primary cosmic rays with the atmosphere) make it possible to study environmental processes such as changes in atmospheric temperature, pressure, or humidity~\citep{Desilets_2010, Kato_2021}. As a major source of radiation exposure for aircraft passengers and crew, astronauts, and spacecraft~\citep{Bottollier-Depois_2012, Chen_2023}, GCRs determine the radiation environment in near-Earth space. Moreover, being the main source of ionization in the lower and middle atmosphere~\citep{Mironova_2015, Singh_2011, Tacza_2024, Yanchukovsky_2026}, cosmic rays influence atmospheric chemical processes, precipitation patterns, and the structure of the global electric circuit. To properly study all these phenomena, information about the shape of GCR energy spectra at different times is required.

GCR fluxes near the Earth are measured using both satellite detectors and ground-based facilities. Certain space experiments (e.g., PAMELA~\citep{Picozza_2007} and AMS-02~\citep{Aguilar_2021}) enable direct measurements of the energy spectra of different GCR species over a wide energy range. However, the acquired data require extensive post-processing, which causes significant delays before the results become available. In addition, on-orbit instruments are subject to rapid radiation degradation with no possibility of repair, and their statistical accuracy is often insufficient to achieve high temporal resolution.

In contrast, ground-based detectors~--- most notably neutron monitors (NMs)~\citep{Simpson_2000}~--- provide real-time data with high temporal resolution and can operate almost continuously for many decades. The disadvantage is that ground-based instruments detect the integral flux of secondary atmospheric particles and therefore do not directly measure GCR spectra. However, in terms of GCR detection, different geographical locations on the Earth's surface can be characterized by effective vertical geomagnetic cutoff rigidity $R_c$ (the minimum rigidity that a particle should have in order to reach a given point while moving from a vertical direction)~\citep{Cooke_1991} and atmospheric depth $h$. NMs situated under different geomagnetic and atmospheric conditions are sensitive to primary particles with different energies. This makes it possible to extract spectral variation information from NM data and to use the global NM network as a unified GCR spectrometer.

Studies of GCR flux variations based on NM data typically rely on the concepts of the NM yield function (YF)~\citep{Mishev_2020} and the coupling function~\citep{Belov_2018, Kovalev_2022}, which enable calculations of the detector’s response to a given flux of primary particles or to relative flux variations, respectively. In such analyses, preliminary assumptions about the energy dependence of the variation amplitude or the shape of the GCR spectrum are made. The corresponding spectral parameters are then optimized so that the theoretically predicted count rate variations for several NMs are consistent with experimental data~\citep{Kobelev_2013, Vais_2023, Vais_2025}.

The drawbacks of this approach are as follows. First, the results depend on the chosen spectral parameterization. Second, there are multiple versions of NM yield and coupling functions~\citep{Debrunner_1982, Nagashima_1989, Сlem_2000, Matthia_2009, Caballero_2012, Maurin_2015, Mangeard_2016, Belov_2018, Mishev_2020}, and their accuracy remains uncertain~~\citep{Koldobskiy_2019}. Third, due to local surroundings, imperfections in the detector electronics, and gradual degradation over time, the yield and coupling functions of real NMs inevitably deviate from those of an ideal instrument. To account for this, empirically determined calibration coefficients are used to link the measured NM count rate with the theoretical prediction~\citep{Ghelfi_2017,Vais_2023}. However, these coefficients cannot fully resolve the issue, as they act merely as scaling factors and do not reflect changes in the shape of the yield or coupling functions. Consequently, the coefficients themselves may vary with solar activity and over different timescales~\citep{Vais_2023}.

As an alternative, one can reconstruct GCR fluxes in specific energy ranges independently at each moment in time, without predefined spectral parameterizations or prior knowledge of detector responses. For this purpose, mathematical models trained on NM data and satellite measurements can be employed. Although GCR fluxes reconstructed by these methods are less reliable from physical perspective, their accuracy is not affected by the uncertainties discussed above. Machine learning techniques have demonstrated their efficiency in studying cosmic rays, solar activity, and their relation to terrestrial processes~\citep{Voyant_2017, Tsai_2022, Malinovic_2023, Belen_2024, Du_2025}. However, to our knowledge, there  are only three papers that utilize machine learning techniques to link ground-based  measurements and dynamics of GCR spectra~\citep{Zhao_2026, Nguyen_2025, Shcherbakova_2026}. All these studies use daily GCR spectra measured by AMS-02 in 2011--2019~\citep{Aguilar_2021-1, Aguilar_2022} as primary source of response variables.

The authors of~\citep{Nguyen_2025} use machine learning methods to reconstruct galactic proton spectra utilizing different indexes representing solar activity and space weather conditions as inputs, and one of those parameters is a count rate of Oulu NM. Therefore, their method is mostly based on correlation between different space weather manifestations rather than physically determined relations between GCR flux and count rates of GCR detectors. Several machine-learning-based models are trained, and their performance is compared with the results obtained in~\citep{Vais_2023} via traditional methods.
In~\citep{Zhao_2026}, count rates provided by worldwide NM network are utilized as the only set of input parameters. Using sophisticated techniques of data imputation, the authors fill the gaps in time series, then train convolutional neural network~\citep{Li_2021} and extend galactic proton time series to the future, until 2024.
Finally, article~\citep{Shcherbakova_2026} is dedicated to reconstruction of helium to proton flux ratio in the past based on NMs data. In addition to daily AMS-02 fluxes, the authors utilize upsampled PAMELA data, train a linear model using stochastic gradient descent~\citep{Amari_1993} and compare reconstruction results with the data from several balloon experiments.

In this paper we use NM data to reconstruct GCR energy spectra taking two different approaches: one based on the NM yield function coupled with a force-field approximation, and another utilizing artificial neural networks.
In comparison to~\citep{Vais_2023, Vais_2025}, this study: 1)~makes use of refined data concerning GCR elemental composition; 2)~tries to fix the flaws of existing NM calibration scheme by using additional parameters; 3)~utilizes a modified version  of the force-field model with two free parameter; 4)~aims to reconstruct and analyze not only proton, but helium fluxes also. Regarding the results and methods presented in~\citep{Nguyen_2025, Zhao_2026, Shcherbakova_2026}, this paper aggregate and broaden them. Specifically, here we: 1)~use several types of machine learning algorithms and provide their result with a physical interpretation; 2)~in addition to NMs count rates, consider several indexes representing solar and geomagnetic activity with physically motivated time delays; 3)~reconstruct both proton and helium fluxes; 4)~extend reconstructed time series to both past and future and verify them not only with AMS-02 measurement results, but using PAMELA data also. Additionally, having GCR time series reconstructed with two methods, we follow~\citep{Nguyen_2025} and compare them to each other. All the topics mentioned in this paragraph are discussed in detail in the corresponding parts of the article.

\section{Data and Methods}
\subsection{Reconstruction of GCR spectra using the NM yield function \label{YF}}

\subsubsection{Contribution of GCR nuclei to the NM count rate}

The theoretical count rate of $i$-th NM $F_i^{\mathrm{theor}}(t)$ at time $t$ can be computed as
\begin{equation}
    F_i^{\mathrm{theor}}(t) = \sum_j \int_{T_c^{i,j}(t)}^{\infty} Y_j^i(T,t)\, A_j\, J_j(T,t)\, dT,
    \label{eq:base}
\end{equation}
where $T_c^{i,j}(t) = \sqrt{\left(\frac{Z_j R_c^i(t)}{A_j}\right)^2 + m^2} - m$ is the minimum kinetic energy per nucleon of primary particles of type~$j$ (characterized by charge number~$Z_j$ and mass number~$A_j$) needed to reach the NM location with effective vertical geomagnetic cutoff rigidity $R_c^i(t)$; $m = 0.938$~GeV; $Y^i_j(T,t)$ is the yield function of the $i$-th NM for one nucleon of GCR species $j$, whose near-Earth energy spectrum is $J_j(T,t)$.

$R_c^i(t)$ gradually changes over time~\citep{Kruchinin_2025} and can vary rapidly during geomagnetic storms (see~\citep{Tyasto_2013} and references therein), but on the time scale of a decade it is usually assumed to remain constant. In turn, $Y^i_j(T,t)$ depends on the amount of matter (the~atmospheric depth) above the detector, $h$, which exhibits short-term variability, and may also change due to shifts in detector properties (e.g., degradation of counter tubes over time or intentional modifications of the detector and its surroundings) or seasonal variations in the local environment. The first two factors are usually accounted for by NM operators, who provide data already corrected for pressure and efficiency; NMs with large seasonal variations can be excluded from analyses. Hence, the time dependence of an NM count rate is determined by variations in the GCR energy spectra associated with solar modulation.

In Equation~\ref{eq:base}, the summation runs over different species of primary GCRs. Although the YFs for proton and helium fluxes differ~\citep{Mishev_2020}, simulations show a similar development pattern (scaled by the number of nucleons) for atmospheric showers induced by helium and heavier nuclei with the same energy per nucleon~\citep{Mishev_2011, Engel_2011}. Therefore, the NM YF $Y^i_j(T)$ can be taken to be the same for all GCR nuclei with $Z \geq 2$.

According to the Parker transport equation~\citep{Parker_1965}, the characteristics of solar modulation are governed by particle rigidity, charge sign, the $Z/A$ ratio, and the shape of the LIS~\citep{Aguilar_2025, Aguilar_2025-1}. Since all nuclei with $Z \geq 2$ have $Z/A \approx 0.5$, fluxes of particles with the same rigidity (or kinetic energy per nucleon) are expected to experience similar modulation~\citep{Koldobskiy_2019}, and observations confirm this for rigidities exceeding several GV~\citep{Aguilar_2025-1}. Thus, the ratio
\begin{equation}
    S_j(T) = \frac{J_j(T,t)}{J_{\mathrm{He}}(T,t)}
\end{equation}
does not depend on solar activity and therefore remains approximately constant over time.

Considering the above, the total contribution of all GCR nuclei with $Z > 2$ can be effectively accounted for by multiplying the helium flux by the scaling factor
\begin{equation}
    S(T) = 1 + \frac{\sum_{j,\, Z_j > 2} A_j \langle J_j(T) \rangle}{4 \langle J_{\mathrm{He}}(T) \rangle},
    \label{eq:scale}
\end{equation}
where $\langle J_j(T) \rangle$ and $\langle J_{\mathrm{He}}(T) \rangle$ are spectra measured over a long time interval. With this in mind, Equation~\ref{eq:base} can be rewritten as
\begin{eqnarray}
    F_i^{\mathrm{theor}}(t) & = & \int_{T_c^{i,p}(t)}^{\infty} Y_p^i(T)\, J_p(T,t)\, dT \; + \nonumber \\
                            & + & 4\int_{T_c^{i,\mathrm{He}}(t)}^{\infty} Y_{\mathrm{He}}^i(T)\, S(T)\, J_{\mathrm{He}}(T,t)\, dT.
    \label{eq:pHe}
\end{eqnarray}

To compute $S(T)$, the authors of~\citep{Koldobskiy_2019} directly considered GCR nuclei with $Z < 9$, whereas the contribution of heavier nuclei was taken as 1.746 times that of oxygen. In this study, we use AMS-02 time-averaged spectra of all GCR nuclei lighter than Si, as well as S and Fe fluxes~\citep{Aguilar_2021, Aguilar_2021-2, Aguilar_2021-3, Aguilar_2021-4, Aguilar_2023-1}. The combined contribution of Ca, Co and Ni nuclei is taken as 0.24 times that of Si~\citep{Israel_2018}, while the cumulative flux of nucleons of P, Cl, Ar, K, Sc, Ti, V, Cr, and Mn nuclei is taken as 1.05 times the flux of nucleons associated with N~\citep{Boschini_2020}. The contribution of nuclei heavier than Ni is assumed to be negligible.

Another innovation is accounting for the helium isotopic composition. Whereas previous studies assumed that helium consisted entirely of $^4\mathrm{He}$, the present work uses a parameterization of the $^3\mathrm{He}/^4\mathrm{He}$ ratio rigidity dependence reported by the AMS collaboration~\citep{Aguilar_2024}. At relatively low energies, $^3\mathrm{He}$ fraction is approximately 15\% and varies with time, exhibiting $\sim 1$\% amplitude over the solar cycle. Consequently, neglecting $^3\mathrm{He}$ leads to an overestimation of the helium contribution to the NM count rate by approximately 3\%, whereas the temporal variability of the isotopic ratio introduces errors an order of magnitude smaller ($\approx 0.2\%$ at low energies).

The results of the $S(T)$ computation are presented in Table~\ref{table:chem}. Energy bins are chosen so that bins No.~5--30 correspond to the AMS-02 daily helium data~\citep{Aguilar_2022}, lines No.~1--4 correspond to the first four bins in AMS-02 daily proton data~\citep{Aguilar_2021-1}, and the remaining energy intervals match those used for the time-averaged helium spectrum~\citep{Aguilar_2021}. Since the authors of~\citep{Koldobskiy_2019} also used AMS-02 measurements as the primary data source, our uncertainties should be similar to theirs, ranging from 0.02 at low energies to approximately 0.07 at high energies. The numbers in the 5-th column of Table~\ref{table:chem} (where the $^3\mathrm{He}$ contribution is not accounted for) agree with~\citep{Koldobskiy_2019} within uncertainties, whereas the values computed with $^3\mathrm{He}$ included are systematically lower.

\begin{table}
\centering
\caption{ \label{table:chem} Ratio of the total number of nucleons in GCR nuclei to that in GCR He.}
\small
\centering
\begin{tabular}{cccccc}
No. & R & T                    & T                              & $S(T)$          & $S(T)$          \\
    & (GV) & for $\frac{Z}{A}=1$ & for $\frac{Z}{A}=\frac{1}{2}$ & no         & with     \\
    &    & (GeV/nuc)              & (GeV/nuc)                        & $^3\mathrm{He}$ & $^3\mathrm{He}$ \\
\hline
 1 & 1.00--1.16 & 0.43--0.55 & 0.12--0.16 &  --- &  --- \\
 2 & 1.16--1.33 & 0.55--0.69 & 0.16--0.22 &  --- &  --- \\
 3 & 1.33--1.51 & 0.69--0.84 & 0.22--0.27 &  --- &  --- \\
 4 & 1.51--1.71 & 0.84--1.0  & 0.27--0.33 &  --- &  --- \\
 5 & 1.71--1.92 &  1.0--1.2  & 0.33--0.40 &  --- &  --- \\
 6 & 1.92--2.15 &  1.2--1.4  & 0.40--0.49 &  --- &  --- \\
 7 & 2.15--2.40 &  1.4--1.6  & 0.49--0.59 & 1.38 & 1.34 \\
 8 & 2.40--2.67 &  1.6--1.9  & 0.59--0.69 & 1.39 & 1.36 \\
 9 & 2.67--2.97 &  1.9--2.2  & 0.69--0.82 & 1.40 & 1.37 \\
10 & 2.97--3.29 &  2.2--2.5  & 0.82--0.96 & 1.41 & 1.38 \\
11 & 3.29--3.64 &  2.5--2.8  & 0.96--1.1  & 1.42 & 1.38 \\
12 & 3.64--4.02 &  2.8--3.2  &  1.1--1.3  & 1.42 & 1.39 \\
13 & 4.02--4.43 &  3.2--3.6  &  1.3--1.5  & 1.43 & 1.39 \\
14 & 4.43--4.88 &  3.6--4.0  &  1.5--1.7  & 1.43 & 1.40 \\
15 & 4.88--5.37 &  4.0--4.5  &  1.7--1.9  & 1.44 & 1.41 \\
16 & 5.37--5.90 &  4.5--5.0  &  1.9--2.2  & 1.44 & 1.41 \\
17 & 5.90--6.47 &  5.0--5.6  &  2.2--2.4  & 1.45 & 1.42 \\
18 & 6.47--7.09 &  5.6--6.2  &  2.4--2.7  & 1.46 & 1.43 \\
19 & 7.09--7.76 &  6.2--6.9  &  2.7--3.1  & 1.46 & 1.43 \\
20 & 7.76--8.48 &  6.9--7.6  &  3.1--3.4  & 1.46 & 1.44 \\
21 & 8.48--9.26 &  7.6--8.4  &  3.4--3.8  & 1.47 & 1.44 \\
22 & 9.26--10.1 &  8.4--9.2  &  3.8--4.2  & 1.47 & 1.45 \\
23 & 10.1--11.0 &  9.2--10   &  4.2--4.6  & 1.47 & 1.45 \\
24 & 11.0--13.0 &   10--12   &  4.6--5.6  & 1.48 & 1.45 \\
25 & 13.0--16.6 &   12--16   &  5.6--7.4  & 1.48 & 1.46 \\
26 & 16.6--22.8 &   16--22   &  7.4--11   & 1.49 & 1.47 \\
27 & 22.8--33.5 &   22--33   &   11--16   & 1.51 & 1.49 \\
28 & 33.5--48.5 &   33--48   &   16--23   & 1.52 & 1.50 \\
29 & 48.5--69.7 &   48--69   &   23--34   & 1.52 & 1.51 \\
30 & 69.7--100  &   69--99   &   34--49   & 1.52 & 1.51 \\
31 &  100--108  &   99--107  &   49--53   & 1.51 & 1.50 \\
32 &  108--116  &  108--115  &   53--57   & 1.51 & 1.50 \\
33 &  116--125  &  115--124  &   57--62   & 1.52 & 1.50 \\
34 &  125--135  &  124--134  &   62--67   & 1.51 & 1.50 \\
35 &  135--147  &  134--146  &   67--73   & 1.51 & 1.50 \\
36 &  147--160  &  146--159  &   73--79   & 1.51 & 1.50 \\
37 &  160--175  &  159--174  &   79--87   & 1.51 & 1.50 \\
38 &  175--192  &  174--191  &   87--95   & 1.51 & 1.50 \\
39 &  192--211  &  191--210  &   95--105  & 1.51 & 1.50 \\
40 &  211--233  &  210--232  &  105--116  & 1.51 & 1.50 \\
41 &  233--259  &  232--258  &  116--129  & 1.52 & 1.51 \\
42 &  259--291  &  258--290  &  129--145  & 1.52 & 1.51 \\
43 &  291--330  &  290--329  &  145--164  & 1.51 & 1.50 \\
44 &  330--379  &  329--378  &  164--189  & 1.51 & 1.50 \\
45 &  379--441  &  378--440  &  189--220  & 1.50 & 1.49 \\
46 &  441--525  &  440--524  &  220--262  & 1.50 & 1.49 \\
47 &  525--643  &  524--642  &  262--321  & 1.50 & 1.49 \\
48 &  643--822  &  642--821  &  321--410  & 1.50 & 1.49 \\
49 &  822--1130 &  821--1129 &  410--564  & 1.50 & 1.49 \\
50 & 1130--1800 & 1129--1799 &  564--899  & 1.50 & 1.49 \\
\end{tabular}
\end{table}

\subsubsection{Calibration of the NM yield function using AMS-02 data}

One can see that Equation~\ref{eq:pHe} includes YFs $Y^i_p(T)$ and $Y^i_{\mathrm{He}}$ specific for each NM. There are different types of NMs, such as IGY~\citep{Simpson_2000}, NM64~\citep{Hatton_1964}, or mini-NM~\citep{Kruger_2015}, and NMs of the same type can have different numbers of counters (standard modules). The YFs of an ideal NM can be described as
\begin{eqnarray}
    Y_{p,\, \mathrm{ideal}}^{i}(T)          & = & N_{\mathrm{mod}}^{i}\, Y_p(T, h_i), \nonumber \\
    Y_{\mathrm{He},\, \mathrm{ideal}}^{i}(T) & = & N_{\mathrm{mod}}^{i}\, Y_{\mathrm{He}}(T, h_i),
    \label{eq:ideal}
\end{eqnarray}
where $N_{\mathrm{mod}}$ is the number of standard modules composing the NM, $h_i$ is the reference atmospheric depth at the NM location, and $Y_p(T, h_i),\ Y_{\mathrm{He}}(T, h_i)$ are theoretically determined YFs to all NMs of that type. In this study, we use data of NM64-type monitors and adopt the YFs model presented in~\citep{Mishev_2020}.

To account for the non-ideality of each specific NM, Equation~\ref{eq:ideal} is modified using empirically determined calibration coefficients. It is common to use a scaling factor $\kappa$ for this purpose~\citep{Ghelfi_2017, Vais_2023}. However, this factor only accounts for a reduced overall detector efficiency and does not capture its energy dependence. Here we address this limitation by introducing an additional parameter $h^*$. Our hypothesis is that the local surroundings of the detector can be interpreted as additional matter that particles must traverse. Consequently, instead of the reference atmospheric depth $h$, one should use $h^*$, which effectively includes the surrounding matter and is therefore somewhat larger:
\begin{eqnarray}
    Y_{p}^{i}(T)           & = & \frac{N_{\mathrm{mod}}^{i}}{\kappa_i}\, Y_{p}(T, h_i^*), \nonumber \\
    Y_{\mathrm{He}}^{i}(T) & = & \frac{N_{\mathrm{mod}}^{i}}{\kappa_i}\, Y_{\mathrm{He}}(T, h_i^*),
    \label{eq:real}
\end{eqnarray}

Therefore, NM count rate can be computed as
\begin{eqnarray}
    & F_i^{\mathrm{theor}}(t; \kappa_i,\!h_i^*)\!=\!\cfrac{N_{\mathrm{mod}}^{i}}{\kappa_i}\!\left( \int_{T_c^{i,p}(t)}^{\infty}\!Y_p(T,h_i^*)J_p(T,t) dT \right.+\nonumber \\
    &                                          + \left. 4 \int_{T_c^{i,\mathrm{He}}(t)}^{\infty} Y_{\mathrm{He}}(T,h_i^*)\, S(T)\, J_{\mathrm{He}}(T,t)\, dT \right).
    \label{eq:for_yf}
\end{eqnarray}

To calibrate the NM YFs, we use daily proton~\citep{Aguilar_2021-1} and helium~\citep{Aguilar_2022} fluxes measured by AMS-02 in 2011--2019. Particles with rigidities greater than 100~GV are accounted for using time-averaged energy spectra~\citep{Aguilar_2021}, and the contribution of particles with $T > 1$~TeV is assumed to be negligible. Numerical integration is performed directly over the energy bins present in the experimental data, without interpolation. For each NM, we randomly select 30\% of the dataset ($K = 772$ points) and fit the theoretically computed count rates to the experimental values:
\begin{equation}
    \sum_{k=1}^{K} \left( F_i^{\mathrm{exp}}(t_k) - F_i^{\mathrm{theor}}(t_k; \kappa_i, h_i^*) \right)^2 \longrightarrow \min ,
\end{equation}
obtaining the optimal values of $\kappa_i$ and $h_i^*$ that define $Y_p^i(T)$ and $Y_{\mathrm{He}}^i(T)$. These values are then fixed and do not vary in the course of GCR reconstruction process described in Section~\ref{sec:ffa}.

This procedure is applied to five NMs, namely BRBG, THUL, APTY, TXBY, and OULU, whose characteristics are presented in Table~\ref{table:nm1}. Mean atmospheric depths at the NM locations are computed using the MATLAB Aerospace Toolbox~\footnote{\url{https://www.mathworks.com/products/aerospace-toolbox.html}}, while effective vertical geomagnetic cutoff rigidities are taken from~\citep{Mishev_2020}. The calculation results are presented in the same table. We find that $\kappa$ values obtained here are similar to those in~\citep{Usoskin_2017, Koldobskiy_2019, Mishev_2020, Vais_2023}. The small discrepancies between the results presented in different papers are most likely caused by distinct ways to consider GCR elemental composition and different datasets utilized to compute these calibration constants. The fitted $h^*$ values are nearly identical to the actual atmospheric depths $h$. We conclude that this parameter is unnecessary, and another approach is needed to account for NM non-ideality.

\begin{table}[ht]
\centering
\caption{\label{table:nm1} Characteristics of neutron monitors used in the YF-based GCR reconstruction.}
\begin{tabular}{ccccccc}
No. & Name & $N_{\mathrm{mod}}$ & $R_c$  & $h$  & $h^*$  & $\kappa$ \\
& & & (GV) & (g/cm$^2$) & (g/cm$^2$) & \\
\hline
1 & BRBG & 3   & 0.00 & 1013 & 1013 & 1.276 \\
2 & THUL & 3   & 0.30 & 1020 & 1020 & 1.622 \\
3 & APTY & 3   & 0.45 & 1002 & 1002 & 1.776 \\
4 & TXBY & 3   & 0.48 & 1024 & 1025 & 1.966 \\
5 & OULU & 1.5 & 0.62 & 1025 & 1025 & 0.898 \\
\end{tabular}
\end{table}

\subsubsection{Force-field approach\label{sec:ffa}}

To reconstruct GCR fluxes with NM YFs, we parameterize their energy spectra as
\begin{equation}
    J_j(T,t) \longrightarrow J_j(T, \bm{\xi}_j(t))
\end{equation}
using a set of time-dependent parameters $\bm{\xi}_j(t)$ specific to each GCR species. The theoretical count rate of an NM with a known YF can then be computed as
\begin{eqnarray}
    & F_i^{\mathrm{theor}} (\bm{\xi}_p(t), \bm{\xi}_{\mathrm{He}}(t)) = \nonumber \\
    & = \frac{N_{\mathrm{mod}}^{i}}{\kappa_i} \left( \int_{T_c^{i,p}(t)}^{\infty} Y_p(T,h_i)\, J_p(T,\bm{\xi}_p(t))\, dT\; + \right. \nonumber \\
    & + 4 \left. \int_{T_c^{i,\mathrm{He}}}^{\infty} Y_{\mathrm{He}}(T,h_i)\, S(T)\, J_{\mathrm{He}}(T,\bm{\xi}_{\mathrm{He}}(t))\, dT \right),
    \label{eq:for_spectra}
\end{eqnarray}
and the best-fit parameter values for each day $t_k$ are determined from
\begin{equation}
    \sum_{i=1}^N \left( F_i^{\mathrm{exp}}(t_k) - F_i^{\mathrm{theor}}(\bm{\xi}_p(t_k), \bm{\xi}_{\mathrm{He}}(t_k)) \right)^2 \rightarrow \min,
\end{equation}
where $N = 5$ is the number of NMs considered. These values fully define the GCR spectra $J_p(T,t)$ and $J_{\mathrm{He}}(T,t)$.

GCR spectra near the Earth can be parameterized using the force-field approach (FFA)~\citep{Gleeson_1968, Caballero-Lopez_2004}:
\begin{eqnarray}
    & J_j(T,t)=\cfrac{(T + m)^2 - m^2}{(T\!+\!m\!+\!\Phi_j(T,t))^2\!-\!m^2}\, J_j^{\mathrm{LIS}}(T\!+\!\Phi_j(T,t)), \nonumber \\
    & \Phi_j(T,t) = \cfrac{|Z_j e|}{A_j}\, \varphi_j(T,t),
    \label{eq:FFApar}
\end{eqnarray}
where $\varphi_j(T,t)$ is the solar modulation potential and $J_j^{\mathrm{LIS}}(T)$ is the local interstellar spectrum. Among many LIS models~\citep{Burger_2000, Vos_2015, Raath_2016, Bisschoff_2019, Aguilar_2021, Shen_2021, Zhu_2024, Shen_2025}, we use those provided in~\citep{Bisschoff_2016}:
\begin{eqnarray}
    & J_p^{\mathrm{LIS}}(T) = 3719 \cfrac{1}{\beta^2} T^{1.03} \left( \cfrac{T^{1.21} + 0.77^{1.21}}{1 + 0.77^{1.21}} \right)^{-3.18}, \nonumber\\
    & J_{\mathrm{He}}^{\mathrm{LIS}}(T) = 195.4 \cfrac{1}{\beta^2} T^{1.02} \left( \cfrac{T^{1.19} + 0.6^{1.19}}{1 + 0.6^{1.19}} \right)^{-3.15},
\end{eqnarray}
where $1 / \beta = (T + m) / \sqrt{T(T + 2m)}$, $T$ is in GeV/nuc, and $J(T)$ is given in $(\mathrm{m}^2\, \mathrm{s}\, \mathrm{sr}\, \mathrm{GeV/nuc})^{-1}$.

In the general case, the solar modulation potential depends on both particle type and energy. However, it is often assumed to be identical across GCR species and energy independent, so a single parameter $\varphi$ describes the shape of the GCR spectra at any given time. Using NM data and Equations~\ref{eq:for_spectra}--\ref{eq:FFApar}, one can reconstruct $\varphi(t)$ and thus the proton and helium fluxes~\citep{Ghelfi_2017, Vais_2023, Vais_2025}.

Although the simplest FFA variant is widely used, it has been shown that an energy-dependent solar modulation potential is required to describe GCR spectra more accurately~\citep{Gieseler_2017, Koldobskiy_2023, Siruk_2024}. In this case, $\varphi$ becomes a function of particle energy (or rigidity) with several parameters to be determined for each $t_k$. Consequently, several FFA variants have been developed~\citep{Cholis_2016, Corti_2016, Gieseler_2017, Shen_2021, Zhu_2025, Shen_2025}. In most of these models, $\varphi(T)$ is characterized by parameters representing high- and low-energy modulation levels, often denoted $\varphi_{\mathrm{high}}$ and $\varphi_{\mathrm{low}}$, together with a transition function. However, since we reconstruct parameters from NM data~--- and NMs are mostly sensitive to primary particles with energies of order 10~GeV~\citep{Asvestari_2017}~--- the sensitivity is dominated by $\varphi_{\mathrm{high}}$, whereas $\varphi_{\mathrm{low}}$ has little effect on the NM count rate and cannot be reliably constrained. Therefore, we adopt the $\varphi(T)$ parameterization proposed in~\citep{Shen_2021}:
\begin{equation}
    \varphi(T,t) = \frac{\varphi_0(t)}{\beta} {T}^{\varphi_1(t)} \left( 1 + \left( \frac{T}{12.85} \right)^{0.9017} \right)^{-2.514},
    \label{eq:shen}
\end{equation}
where $T$ is in GeV/nuc and $\varphi_0(t)$, $\varphi_1(t)$ are free parameters.
This model is constructed such that both parameters affect the shape of the high-energy part of the GCR spectra and therefore contribute to the NM count rate. Nevertheless, this parameterization is not perfect. Parameters fitted from NM data are appropriate for the multi-GeV range, whereas precise description of particle fluxes at $\sim 100$~MeV requires somewhat different values. This leads to inaccuracies in the low-energy part of the spectra, where solar modulation is stronger and small parameter shifts can cause large discrepancies between predicted and measured fluxes. For this reason, in this section we use only NMs located at places with low $R_c$ (Table~\ref{table:nm1}), which are more sensitive to low-energy GCRs~\citep{Asvestari_2017, Siruk_2023}.

\subsection{Reconstruction of GCR spectra with artificial neural networks \label{ANN}}

\subsubsection{Data preparation \label{preparation}}

To reconstruct GCR spectra using machine learning, the models require training examples of inputs and outputs. As response variables, we use daily proton~\citep{Aguilar_2021-1} and helium~\citep{Aguilar_2022} spectra measured by AMS-02 in 2011--2019, in units of $(\mathrm{m}^{2}\, \mathrm{s}\, \mathrm{sr}\, \mathrm{GeV/nuc})^{-1}$. The predictor variables are: daily averaged count rates (in s$^{-1}$) of seventeen NMs obtained from NMDB~\footnote{\url{https://www.nmdb.eu/nest/}} and IZMIRAN database~\footnote{\url{http://cr0.izmiran.ru/common/links.htm}}, daily sunspot numbers (SSN) provided by WDC-SILSO~\citep{SILSO_Sunspot_Number}, Ap index values provided by the Geomagnetic Observatory Niemegk~\citep{Matzka_2021}, and the heliospheric magnetic field polarity $A$ according to WSO observations of the solar polar magnetic field~\footnote{\url{http://wso.stanford.edu/Polar.html}}.

\begin{table}[h]
\centering
\caption{\label{table:nm} Specifications of neutron monitors used for the neural-network-based GCR reconstruction.}
\begin{tabular}{ccccccc}
No. & Name & lat.  & lon.  & alt.  & h  & $R_c$  \\
& & (deg)& (deg)&(m)&(g/cm$^2$)&(GV)\\
\hline
 1 & BRBG &  78.06 &  14.2  &   70 & 1013 & 0.00 \\
 2 & MRNY & -66.55 &  93.02 &   30 & 1013 & 0.03 \\
 3 & SOPO & -90.0  &    --- & 2820 &  698 & 0.10 \\
 4 & THUL &  76.5  & -68.7  &   26 & 1020 & 0.30 \\
 5 & APTY &  67.57 &  33.4  &  181 & 1002 & 0.45 \\
 6 & TXBY &  71.36 & 128.54 &    0 & 1024 & 0.48 \\
 7 & OULU &  65.05 &  25.47 &   15 & 1025 & 0.62 \\
 8 & KERG & -49.35 &  70.25 &   33 & 1009 & 1.14 \\
 9 & YKTK &  62.01 & 129.43 &  105 & 1015 & 1.37 \\
10 & MOSC &  55.47 &  37.32 &  200 & 1010 & 2.13 \\
11 & NVBK &  54.48 &  83    &  163 & 1015 & 2.40 \\
12 & LMKS &  49.20 &  20.22 & 2634 &  753 & 3.84 \\
13 & JUNG &  46.55 &   7.98 & 3570 &  674 & 4.49 \\
14 & AATB &  43.14 &  76.6  & 3340 &  695 & 5.95 \\
15 & MXCO &  19.33 & -99.18 & 2274 &  803 & 8.28 \\
16 & ATHN &  37.97 &  23.78 &  260 & 1017 & 8.53 \\
17 & PSNM &  18.59 &  98.49 & 2565 &  773 & 16.8 \\
\end{tabular}
\end{table}

Specifications of the neutron monitors selected for this study are presented in Table~\ref{table:nm}. Count rates of some NMs (TXBY, YKTK, LMKS) exhibit jumps not seen in other stations, likely related to changes in detector efficiency or data acquisition systems; these defects are corrected manually. NM data are then preprocessed as follows:
\begin{enumerate}
    \item We compute a 55-day running local median $M(t)$ and a local scaled median absolute deviation (SMAD) of the NM count rate:
    \begin{eqnarray}
        & & M_i(t_k) = \operatorname{median}(F_i(t_m)), \nonumber \\
        & & \mathrm{SMAD}_i(t_k) = \frac{\operatorname{median}(|F_i(t_m) - M_i(t_k)|)}{-\operatorname{erfcinv}(3/2) \cdot \sqrt{2}}, \\
        & & m = k - 27,\ \ldots,\ k + 27, \nonumber
        \label{eq:med}
    \end{eqnarray}
    where $\operatorname{erfcinv}$ is the inverse complementary error function.
    \item We remove data points that deviate by more than three local $\mathrm{SMAD}(t)$ from the local median $M(t)$.
    \item We reconstruct missing data: for short gaps ($\leq 3$~days) by linear interpolation and for longer gaps by substituting $M(t)$ values.
\end{enumerate}

Step 2 is implemented so as not to reject outliers observed simultaneously by three or more NMs, thereby retaining data points corresponding to FDs.

\begin{figure}[h]
\centering
\includegraphics[width=\linewidth]{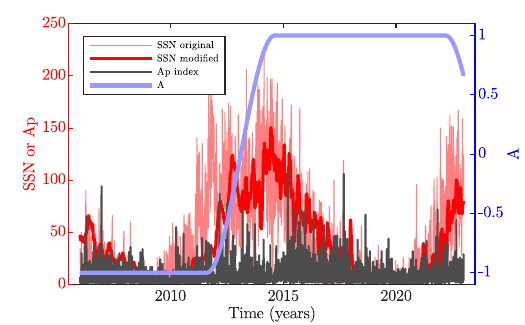}
\caption{\label{fig_indexes} Time series of indices of solar (sunspot number) and geomagnetic (Ap) activity, and the configuration of the heliosphere (heliospheric magnetic field polarity) in 2006--2022.}
\end{figure}

AMS-02 data are processed in a similar way: outliers are removed, and short gaps are filled by linear interpolation. Unlike the NM data, we do not use a running median to fill large gaps, so the response variables retain missing entries.

In addition to NM data, we account for the large-scale heliospheric magnetic field configuration via the polarity~$A$. When the polarity is positive ($A > 0$), the field lines in the northern hemisphere point outward from the Sun; for negative polarity ($A < 0$), their direction is opposite.

Charged particles with charge $q$ experience gradient and curvature drifts in this field. If $qA > 0$, particles preferentially propagate through the heliospheric polar regions; if $qA < 0$, they drift along the heliospheric current sheet, spending more time in transit and losing more energy~\citep{Strauss_2011}. To quantify this effect, we set $A = \pm 1$ during intervals of well-defined positive/negative polarity, respectively. During polarity reversals, $A \in [-1, 1]$ is modeled by a smooth-step function
\begin{equation}
    A(x) = 6x^2 - 4x^3 - 1,
\end{equation}
with $x = 0.5 \pm (t - t_{\mathrm{transition}})/\Delta t$ (the plus sign for a transition from $A < 0$ to $A > 0$, the minus sign for the opposite case), clipped to $[0, 1]$. We take $\Delta t = 3$~years following~\citep{Shen_2021}. The two most recent transitions are taken as March 3, 2013 ($A < 0 \to A > 0$) and September 17, 2023 ($A > 0 \to A < 0$).

We also include the current level of solar activity, represented by the SSN. Information about changes on the solar surface does not propagate through the heliosphere instantaneously but is carried outward by the solar wind with the magnetic field frozen in. Upon reaching the outer heliosphere, this field, affects GCR transport. Accordingly, the SSN input to the model is averaged over the GCR propagation time and delayed by the sum of solar-wind and GCR propagation times:
\begin{eqnarray}
    \mathrm{SSN}_{\pm}(t_k) = \frac{1}{\tau_{\pm}} \sum_{m:\ t_m = t_k - \tau_{\pm}/2}^{m:\ t_m = t_k + \tau_{\pm}/2} \mathrm{SSN}(t_m - \delta t_{\pm}),
\end{eqnarray}
where the subscripts $\pm$ refer to the sign of the product $qA$. Following~\citep{Shen_2021}, during intervals containing a polarity reversal the effective sunspot number is
\begin{equation}
    \mathrm{SSN}_{\mathrm{eff}} = \frac{1 + qA}{2}\, \mathrm{SSN}_{+} + \frac{1 - qA}{2} \mathrm{SSN}_{-}.
\end{equation}

In this study galactic protons and helium nuclei are considered, so in both cases the sign of $qA$ is the same. Based on~\citep{Cholis_2016, Shen_2021, Strauss_2011, Koldobskiy_2022, Tomassetti_2022, Wang_2025}, we set $\tau_{+} = 1$ Bartels rotation (BR) and $\delta t_{+} = 4$~BRs (i.e., 27 and 108 days, respectively), and $\tau_{-} = 3$~BRs and $\delta t_{-} = 13$~BRs. Strictly speaking, propagation time should depend on particles energy~\citep{Strauss_2011, Tomassetti_2022, Wang_2025}, but to not overcomplicate the model we fix it at a value characteristic of particles with an energy of several GeV.

The resulting time series for 2006--2022 are shown in Figure~\ref{fig_indexes}. In addition to SSN and $A$, we include the Ap index, which is not specially preprocessed. This parameter reflects geomagnetic activity, which alters the geomagnetic cutoff rigidity~\citep{Danilova_2025} and may therefore affect NM count rates (see~\citep{Kovalev_2022} and Equation~\ref{eq:base}). These three datasets are used alongside the count rates of the seventeen NMs, yielding twenty predictor variables in total.

\subsubsection{Model training \label{training}}

Before being fed into the models, all datasets described above are standardized: the mean over the full interval is subtracted from each time series and the result is divided by the standard deviation,
\begin{eqnarray}
    & & \mu(T_l) = \frac{1}{K} \sum_{k=1}^{K} J(T_l,t_k), \nonumber \\
    & & \varsigma(T_l) = \sqrt{\frac{1}{K-1} \sum_{k=1}^{K} \left(J(T_l,t_k) - \mu(T_l)\right)^2}, \\
    & & j(T_l,t_k) = \frac{J(T_l,t_k) - \mu(T_l)}{\varsigma(T_l)}, \nonumber
    \label{eq:norm}
\end{eqnarray}
and the two sample statistics $\mu$ and $\varsigma$ are stored to invert the normalization for the results of prediction.

To identify an optimal model, we evaluated several algorithm families: three linear models (SGD regression~\citep{Amari_1993}, least-angle regression~\citep{Efron_2004}, and Lasso~\citep{Tibshirani_1996}), two tree-based methods (random forest~\citep{Breiman_2001} and gradient-boosted trees~\citep{Natekin_2013}), and two types of neural networks (multilayer perceptron, MLP~\citep{Popescu_2009}, and a convolutional neural network, CNN~\citep{Li_2021}).

\begin{figure}
\centering
\includegraphics[width=0.86\linewidth]{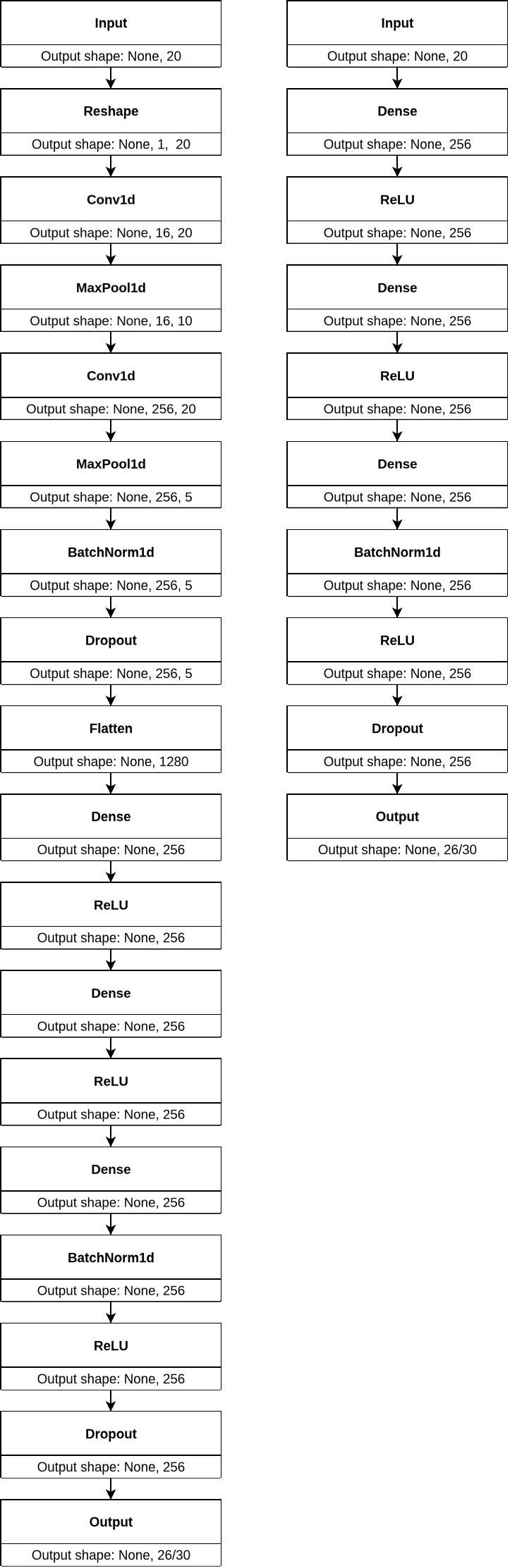}%
\caption{\label{fig_schematic} Schematic for CNN and MLP neural architectures.}
\end{figure}

Model selection uses five-fold cross-validation and grid search for hyperparameter tuning. We found that more traditional selection metrics, such as $R^2$, appear to be very close to their perfect value in too many cases, so it is difficult to choose between algorithms relying on them. Therefore, we wanted our metric to be more punishing towards incorrectly predicted values, so we organized it as follows:
\begin{eqnarray}
  &\eta =\label{eq:eta} \\
  &\frac{1}{K}\!\sum_{k=1}^{K}\!H\!\left( \frac{1}{2}\!-\!\sum_{l=1}^{L}\!H\!\left( \frac{|J^{\mathrm{theor}}(T_l,t_k) - J^{\mathrm{exp}}(T_l,t_k)|}{3\sigma(T_l,t_k)}\!-\!1 \right)\!\right),\nonumber 
\end{eqnarray}
where $H(x)$ is the Heaviside step function. If for at least one of the $L$ energy bins on day $t_k$ the deviation $|J^{\mathrm{theor}}-J^{\mathrm{exp}}|$ exceeds $3\sigma$, the spectrum for that day is considered incorrectly reconstructed; $\eta$ is the fraction of days in the test sample reconstructed correctly by this criterion.

Cross-validation shows that, with optimal hyperparameters, both neural networks reach comparable accuracy for proton and helium fluxes. Linear models perform substantially worse and are discarded. Tree-based models demonstrate precision comparable to that of neural networks, but are rejected at a later stage for reasons given in Section~\ref{2006}. Consequently, we select two neural architectures (CNN and MLP) as our main machine-learning-based models and detail them below.

Neural network architecture schematics are presented in Figure~\ref{fig_schematic}. Both CNN and MLP are unified models with 20 inputs and 26 or 30 outputs for helium or proton data, respectively. Training minimizes the mean squared error over all energy bins and days,
\begin{equation}
    \frac{1}{KL} \sum_{k=1}^{K} \sum_{l=1}^{L} \left( {j^{\mathrm{exp}}(T_l,t_k)} - j^{\mathrm{theor}}(T_l,t_k;\,\bm{w}) \right)^2 \longrightarrow \min,
\end{equation}
where $j(T,t)$ denotes the normalized flux (Equation~\ref{eq:norm}) and $\bm{w}$ are the network parameters.

Optimized hyperparameters are as follows. For the CNN: two dropout rates of 0.1 and 0.5 (regularization by randomly disabling neurons), learning rate 0.005, and a step learning-rate scheduler with step size of 5 epochs. For the MLP: three hidden layers with 256 neurons each; batch normalization and dropout of 0.5 on the third layer; learning rate 0.01. We also tested learning-rate decay via a scheduler parameter $\gamma$, but $\gamma = 1$ was selected, yielding a constant learning rate for both models. Both networks use the RAdam optimizer~\citep{Liu_2019}.

After hyperparameter tuning, the dataset is split into training (70\%) and test (30\%) subsets, and the final models are trained.  This yields four networks: one CNN and one MLP for each species (protons and helium). The test subset is utilized to calculate metrics representing models performance (see Section~\ref{results1}), specifically $\eta$ (Equation~\ref{eq:eta}) and mean absolute percentage error (Equation~\ref{eq:mape}).

\begin{figure}
\centering
\begin{subfigure}[b]{0.48\linewidth}
\includegraphics[width=\linewidth, height=1.272\textwidth]{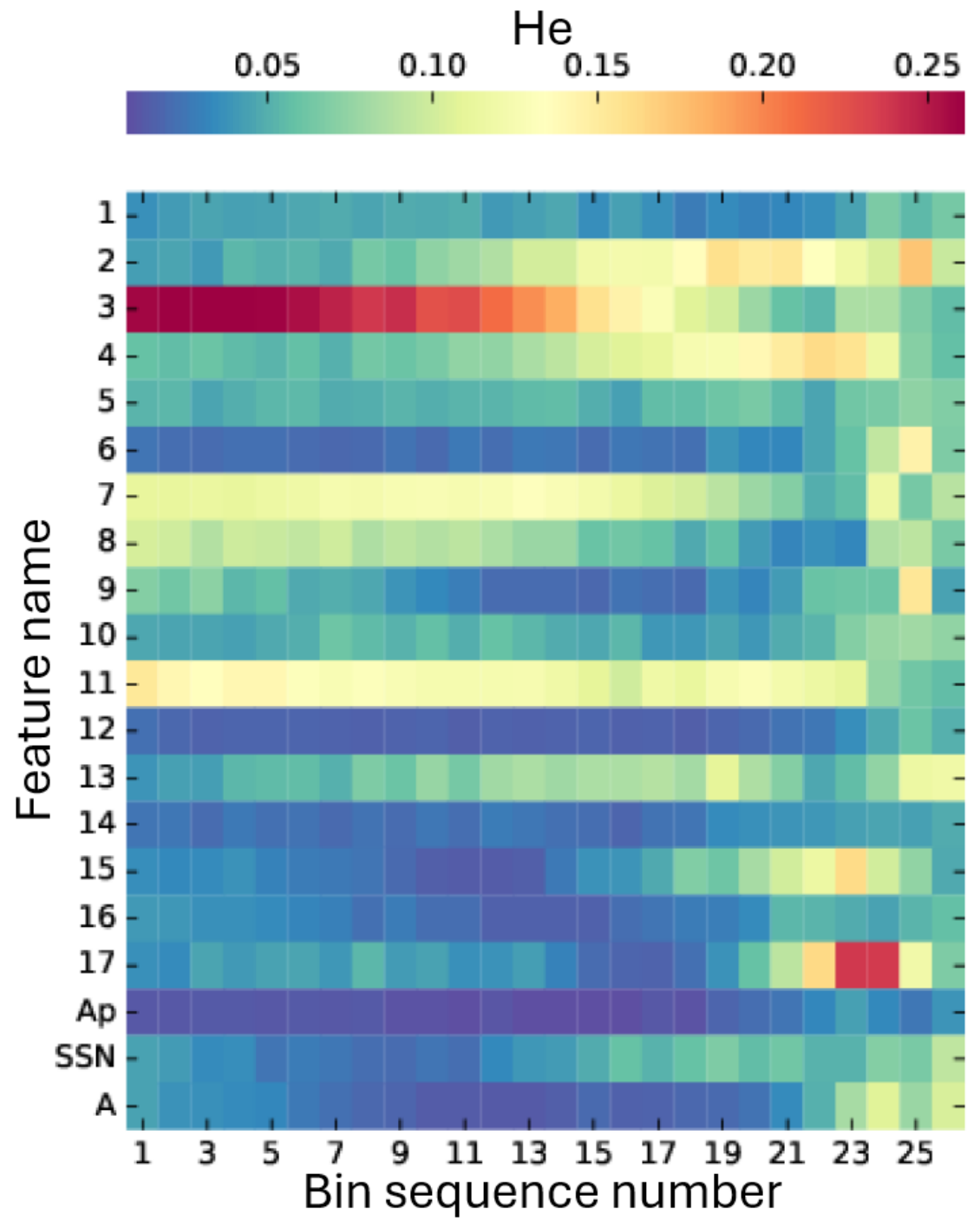}%
   \caption{\label{fig_shap1}}
\end{subfigure}
\begin{subfigure}[b]{0.48\linewidth}
\includegraphics[width=\linewidth, height=1.272\linewidth]{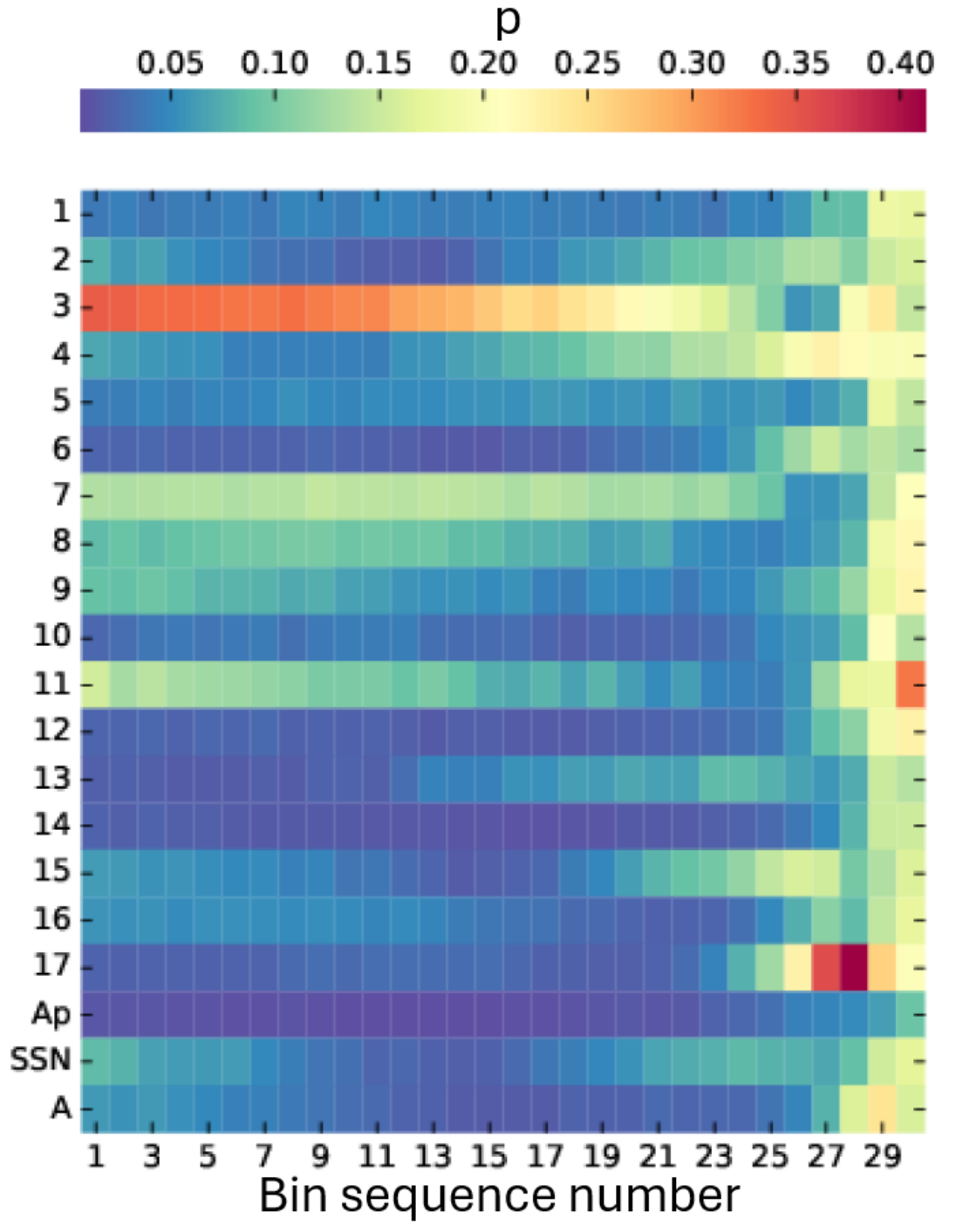}
   \caption{\label{fig_shap2}}
\end{subfigure}
\caption{\label{fig_shap} Mean absolute SHAP values for the MLP models predicting proton (a) and helium (b) fluxes.}
\end{figure}

To interpret the models, we use SHAP~\citep{Lundberg_2017} to estimate the contribution of each predictor to each target component. Figures~\ref{fig_shap1} and~\ref{fig_shap2}  show examples for the two MLP models. The vertical axis lists NM indices as in Table~\ref{table:nm}; the horizontal axes shows energy-bin indices for the reconstructed proton and helium spectra (see Table~\ref{table:chem}, rows No.~1--30 for protons and No.~5--30 for helium).

A neutron monitor at geomagnetic cutoff rigidity $R_c$ does not detect particles below that rigidity; thus, the primary energy to which a station is most sensitive increases with $R_c$~\citep{Asvestari_2017}. An exception is station No.~3 (SOPO), which combines low $R_c$ with high altitude, making it exceptionally sensitive to low-energy particles~\citep{Poluianov_2017, Poluianov_2022} and hence to solar modulation.

Since Table~\ref{table:nm} is ordered by increasing $R_c$, one expects earlier stations to contribute mainly to low-energy predictions, while later stations matter at higher energies. The SHAP results in Figure~\ref{fig_shap} agree with this: SOPO contributes most below $\sim 5$~GV, whereas PSNM (No.~17) becomes influential only above $\sim 10$~GV and is decisive between 20 and 70 GV.

Regarding solar and geomagnetic activity predictors, the SSN and the heliospheric magnetic field polarity $A$ are informative for GCR flux prediction. By contrast, the Ap index should matter only during strong geomagnetic disturbances due to their influence on geomagnetic cutoff rigidity in NMs locations and therefore NMs count rate~\citep{Tyasto_2013, Danilova_2025, Wang_2023}. However, these events are rare relative to the dataset duration: in 2011-2019, there are only 20 days with average Ap index greater than 48 (Kp index greater than 5), and for some of those days AMS-02 data are missing. Therefore, for most instances, Ap index relationship to GCR flux is weak, and the learning algorithm downweights this feature during training. For the same reason, SHAP values are low in the majority of cases, so their averages demonstrated in Figure~\ref{fig_shap} are also close to zero. The question of models performance during intense geomagnetic storms and strong FDs requires a separate detailed investigation and will be addressed in the future. 

SHAP results for the CNN are similar to those for the MLP, indicating that both models behave consistently with physical expectations.

\section{Results and Discussion}
\subsection{Methods performance in 2011--2019 \label{results1}}

\begin{figure*}
\centering
\begin{subfigure}{\linewidth}
\includegraphics[width=\linewidth]{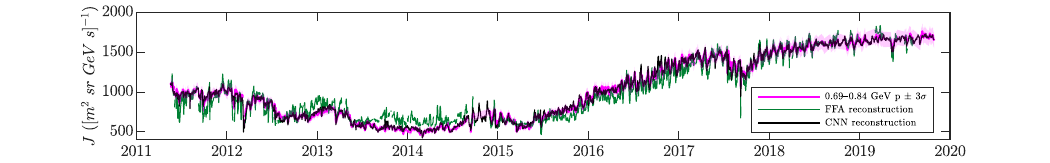}
\caption{\label{fig_Mid_p_low}}
\end{subfigure}
\begin{subfigure}{\linewidth} 
\includegraphics[width=\linewidth]{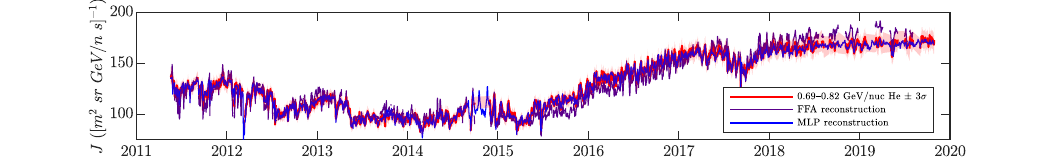}
\caption{\label{fig_Mid_he_low}}
\end{subfigure}
\begin{subfigure}{\linewidth} 
\includegraphics[width=\linewidth]{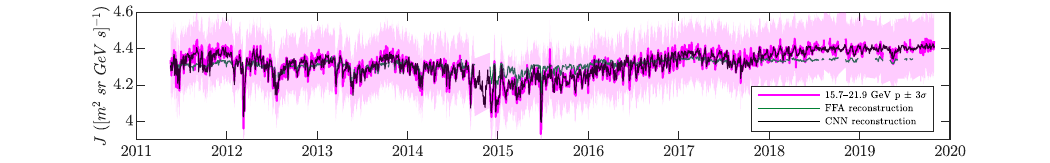}
\caption{\label{fig_Mid_p_high}}
\end{subfigure}
\begin{subfigure}{\linewidth} 
\includegraphics[width=\linewidth]{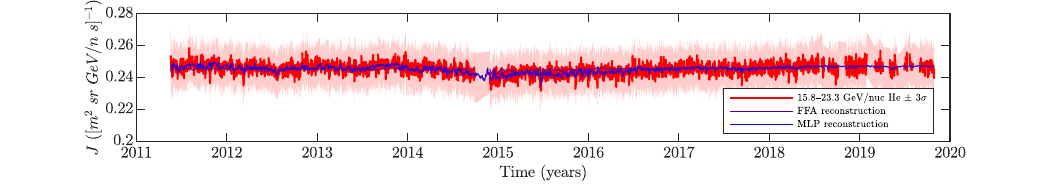}
\caption{\label{fig_Mid_he_high}}
\end{subfigure}
\caption{\label{fig_Mid_time_series} Comparison of AMS-02 daily data (2011--2019) and the results of GCR reconstruction with two methods. Panel (a): low-energy proton flux; panel (b): low-energy helium flux; panels (c) and (d):  high-energy  proton and helium fluxes, respectively.}
\end{figure*}

Using the FFA-based approach described earlier, we reconstruct daily proton and helium fluxes for 2011--2019 from NM data. Since the reconstruction quality is expected to be best around $T \approx 10$~GeV/nuc, we assess its performance in the low-energy range (where solar modulation parameters may differ from those inferred here) and at relatively high energies (where modulation effects are weak and comparable to statistical fluctuations). As an example, the top panels of Figure~\ref{fig_Mid_time_series} show time series of low-energy proton and helium fluxes alongside AMS-02 data, while the bottom panels present high-energy fluxes.

In the low-energy range, the reconstruction is generally good. However, systematic discrepancies appear during certain periods, such as the solar activity maximum for protons, the solar minimum for helium, and the transitional phase of the solar cycle (2015--2016) for both species. Additionally, the FFA predicts amplitudes of short-term changes (both recurrent variations and FDs) to be substantially larger than observed.

For high energies, the FFA-based reconstruction underestimates solar modulation effects on both long and short timescales. Specifically, the helium flux at $T \approx 20$~GeV/nuc is predicted to remain nearly constant, whereas the AMS-02 data clearly exhibit long-term variations on top of statistical fluctuations.

\begin{figure*}
\centering
\begin{subfigure}{\linewidth}
\includegraphics[width=\linewidth]{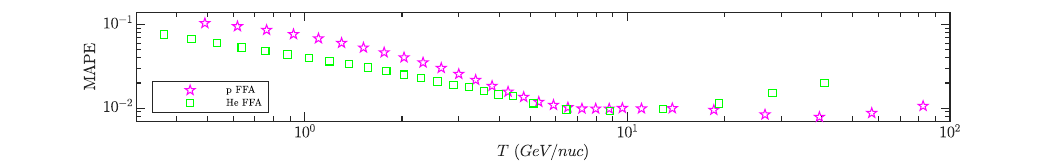}
\caption{\label{fig_ffm_mape}} 
\end{subfigure}
\begin{subfigure}{\linewidth}
\includegraphics[width=\linewidth]{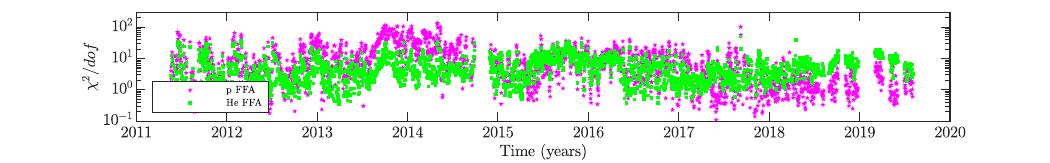}
\caption{\label{fig_ffm_chi2}} 
\end{subfigure}
\begin{subfigure}{\linewidth}
   \includegraphics[width=\linewidth]{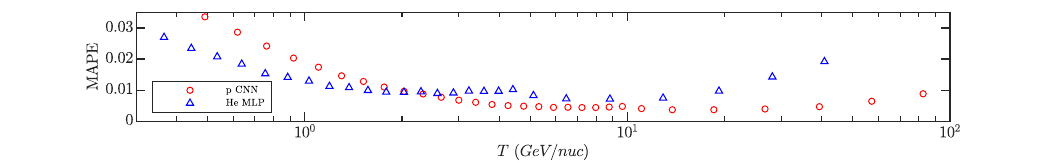}
\caption{\label{fig_networks_mape}}
\end{subfigure}
\begin{subfigure}{\linewidth}
   \includegraphics[width=\linewidth]{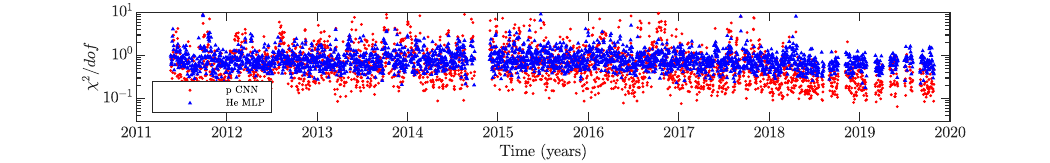}
\caption{\label{fig_networks_chi2}}
\end{subfigure}
\caption{\label{fig_characteristics} Panels (a) and (b): mean absolute percentage error of FFA-based reconstruction of proton and helium fluxes as a function of particle energy, as well as $\chi^2/\mathrm{dof}$ as a function of time, respectively. Panels (c) and (d): same for the CNN-based proton flux  and MLP-based helium flux reconstruction.}
\end{figure*}

To quantify the time–energy dependence of the reconstruction quality, we compute two metrics: the mean absolute percentage error (MAPE) for each energy bin,
\begin{equation}
    \mathrm{MAPE}(T_l) = \frac{1}{K} \sum_{k=1}^{K} \frac{|J^{\mathrm{theor}}(T_l,t_k) - J^{\mathrm{exp}}(T_l,t_k)|}{J^{\mathrm{exp}}(T_l,t_k)},
    \label{eq:mape}
\end{equation}
and the chi-squared per degree of freedom at every moment in time,
\begin{equation}
    \chi^2/\mathrm{dof}(t_k) = \frac{1}{L-2} \sum_{l=1}^{L} \frac{\left(J^{\mathrm{theor}}(T_l,t_k) - J^{\mathrm{exp}}(T_l,t_k)\right)^2}{\sigma(T_l,t_k)^2},
    \label{eq:chi2}
\end{equation}
where $\sigma^2$ is the sum of squared statistical and total systematic uncertainties of the experimental data, $K = 2572$ is the number of days in the time series, and $L$ is the number of energy bins in AMS-02 daily proton ($L = 30$) and helium ($L = 26$)  spectra.

The results are shown in Figures~\ref{fig_ffm_mape} and~\ref{fig_ffm_chi2}. MAPE is plotted versus the kinetic energy per nucleon; the corresponding rigidity boundaries are listed in Table~\ref{table:chem} (rows No.~1--30 for protons and No.~5--30 for helium). The reconstruction quality improves with particle energy, ranging from approximately 10\% at the lowest energies to $\approx 1$\% for $T = 6$--20~GeV/nuc, and then degrades due to the statistical limitations of the AMS-02 data (see Figure~\ref{fig_Mid_he_high}).
As for $\chi^2/\mathrm{dof}$ time series, both exhibit long-term variations correlated with the solar cycle, which are more pronounced for proton fluxes. Recurrent variations with a period of approximately one month are also visible, likely reflecting the 27-day GCR variations. This indicates that the reconstructed spectrum periodically matches the experimental spectrum better or worse, implying that the model does not fully capture the energy dependence of GCR variations. Typical $\chi^2/\mathrm{dof}$ values lie within the $10^0$--$10^2$ range. Overall, the reconstruction quality is sufficient for practical purposes, although this approach does not reproduce all features of solar modulation.

The time series of particle fluxes at different energies reconstructed with machine-learning-based methods are also shown in Figure~\ref{fig_Mid_time_series}. Since the reconstruction quality is similar across models, we present CNN results for protons and MLP results for helium as examples. The neural-network-based reconstructions agree better with the experimental data than the NM YF-based method. Long-term GCR variations are well reproduced at both low and high energies. Furthermore, the models allow to reconstruct short-term disturbances and seamlessly fill gaps in the AMS-02 data. However, the question of models ability to accurately reproduce amplitudes of periodic variations and FDs needs further exploration.

Figures~\ref{fig_networks_mape} and~\ref{fig_networks_chi2} present the metrics defined by Equations~\ref{eq:mape} and \ref{eq:chi2} for proton and helium fluxes reconstructed with CNN and MLP models, respectively. In case of proton fluxes reconstructed with MLP and helium fluxes reconstructed with CNN, these metrics take similar values. The $L-2$ factor in Equation~\ref{eq:chi2} is due to the presence of two free parameters in modified FFA. Regarding neural networks, their internal free parameters (weights) are adjusted to describe the entire training set, so it is difficult to estimate the number of degrees of freedom for a single day. For consistency, we keep denominator equal to $L-2$ in this case. 

\begin{figure*}
\centering
\begin{minipage}{0.585\textwidth}
    \begin{subfigure}{\textwidth}
\includegraphics[width=\linewidth]{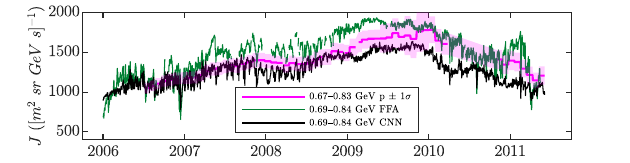}
  \caption{\label{fig_PAMELA_p_low}}
\end{subfigure}
\begin{subfigure}{\textwidth}
\includegraphics[width=\linewidth]{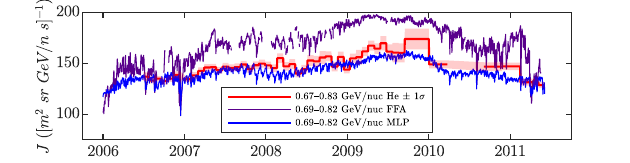}
  \caption{\label{fig_PAMELA_he_low}}
\end{subfigure}
\begin{subfigure}{\textwidth}
\includegraphics[width=\linewidth]{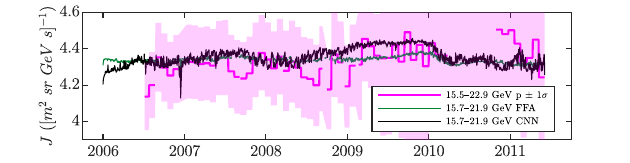}
  \caption{\label{fig_PAMELA_p_high}}
\end{subfigure}
\begin{subfigure}{\textwidth}
\includegraphics[width=\linewidth]{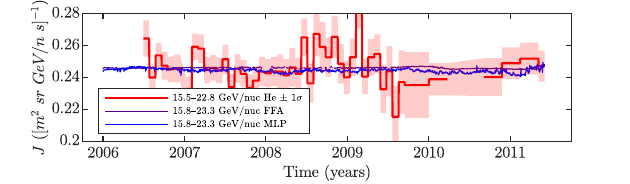}
  \caption{\label{fig_PAMELA_he_high}}
\end{subfigure}
\end{minipage}
\begin{minipage}{0.395\textwidth}
    \begin{subfigure}{\textwidth}
\includegraphics[width=\linewidth]{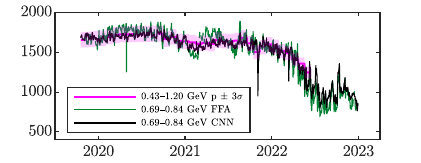}
  \caption{\label{fig_AMS_p_low}}
\end{subfigure}
\begin{subfigure}{\textwidth}
\includegraphics[width=\linewidth]{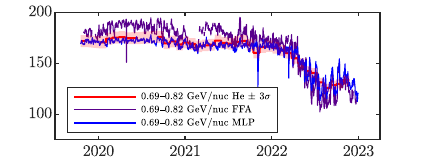}
  \caption{\label{fig_AMS_he_low}}
\end{subfigure}
\begin{subfigure}{\textwidth}
\includegraphics[width=\linewidth]{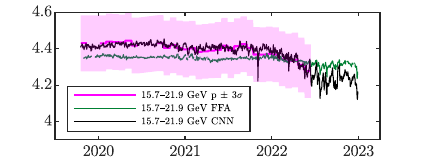}
  \caption{\label{fig_AMS_p_high}}
\end{subfigure}
\begin{subfigure}{\textwidth}
\includegraphics[width=\linewidth]{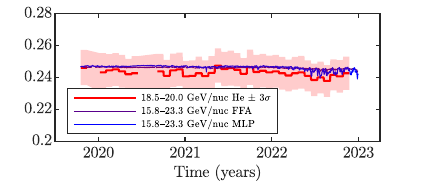}
  \caption{\label{fig_AMS_he_high}}
\end{subfigure}
\end{minipage}

\caption{\label{fig_all_time_series} Panels (a)--(d): comparison of PAMELA  1--3-month-averaged data (2006--2011) and the results of GCR reconstruction with two methods. Panel (a): low-energy proton flux; panel (b): low-energy helium flux; panels (c) and (d):  high-energy  proton and helium fluxes, respectively. Panels (e)--(h): same for AMS-02 BR-averaged data (2019--2022).}
\end{figure*}

Figure~\ref{fig_networks_mape} shows MAPE on the test sample versus the bin index. The dependence is similar to that for the YF-based method: accuracy is worst at the lowest energies and improves with bin index. However, the plateau in reconstruction quality is reached earlier, from $T \approx 3$~GeV/nuc for protons and $T \approx 1$~GeV/nuc for helium nuclei instead of the $T \approx 6$~GeV/nuc in case of the YF-based method (see Figure~\ref{fig_ffm_mape}). Overall, replacing the YF+FFA approach with neural networks reduces MAPE by a factor of approximately 4 at low energies and by about 2 at high energies.

Figure~\ref{fig_networks_chi2} shows $\chi^2/\mathrm{dof}$ for the proton flux for each day in the period considered. For both models, this parameter mostly ranges from 0.1 to 2 and never exceeds 10, indicating a good description of the energy spectrum. A long-term trend associated with the solar cycle is visible in both time series, though less pronounced than for the YF-based method (see Figure~\ref{fig_ffm_chi2}). Thus, by the $\chi^2/\mathrm{dof}$ criterion, neural-network-based reconstructions are roughly an order of magnitude better than those based on the NM YF and FFA.

In addition to MAPE and $\chi^2/\mathrm{dof}$, we compute the $\eta$ metric on the test set (Equation~\ref{eq:eta}) for all machine-learning-based models. For the CNN, $\eta = 0.897$ (protons) and $\eta = 0.902$ (helium); for the MLP, $\eta = 0.907$ for both species. Hence, on data not used for training, the neural networks correctly reconstruct GCR spectra in about 90\% of cases.

The methods described above can be used to reconstruct GCR energy spectra during periods when daily AMS-02 data are unavailable. Particle flux variations after the training and validation interval~--- from late 2019 to the present~--- can be reconstructed directly using the trained models. Reconstructing fluxes in the past (before May 20, 2011) is complicated by the absence of data from certain NMs: first from SOPO (No.~3), then from PSNM (No.~17), which possess heightened sensitivity to the lowest and highest primary energies, respectively. Data from MRNY (No.~2) also become unavailable beyond a certain point. To address this issue, we train two additional pairs of models (one CNN and one MLP for both protons and helium) with reduced input features: removing the three aforementioned NMs leaves 17 features instead of the original 20.

We refer to the original versions as the full models and those used for past reconstructions as the truncated models. Since SOPO and PSNM were important predictors (see Figures~\ref{fig_shap1} and~\ref{fig_shap2}), their absence results in a 1--2\% accuracy loss. 

\subsection{Reconstruction of GCR spectra in 2006--2011 and 2019--2022 \label{2006}}

\begin{figure*}
\centering
\begin{minipage}{0.49\textwidth}
    \begin{subfigure}[b]{\textwidth}
\includegraphics[width=\linewidth]{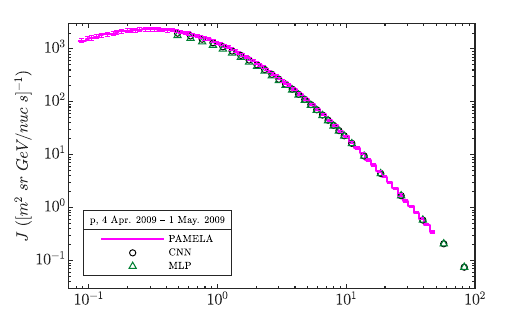}
  \caption{\label{fig_PAMELA_p_spectrum}}
\end{subfigure}
\begin{subfigure}[b]{\textwidth}
\includegraphics[width=\linewidth]{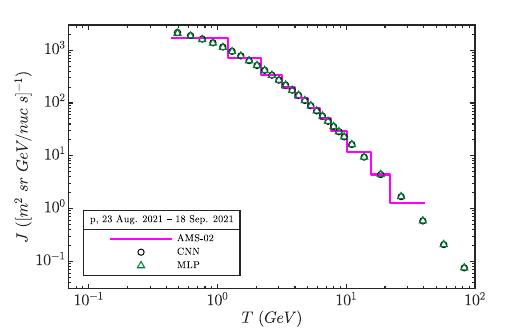}
  \caption{\label{fig_AMS_p_spectrum}}
\end{subfigure}
\end{minipage}
\begin{minipage}{0.49\textwidth}
    \begin{subfigure}[b]{\textwidth}
\includegraphics[width=\linewidth]{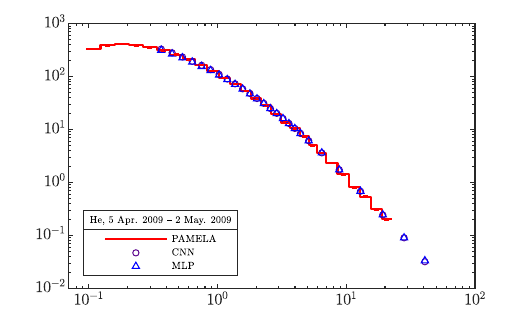}
  \caption{\label{fig_PAMELA_he_spectrum}}
\end{subfigure}
\begin{subfigure}[b]{\textwidth}
\includegraphics[width=\linewidth]{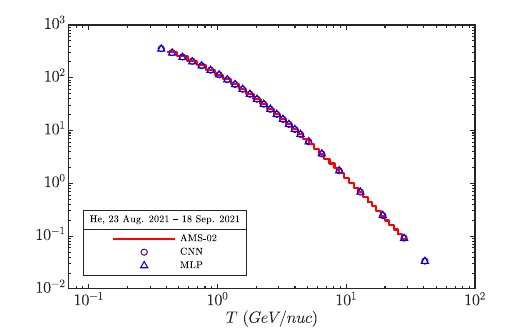}
  \caption{\label{fig_AMS_he_spectrum}}
\end{subfigure}
\end{minipage}
\caption{\label{fig_spectra} Panels (a) and (b): proton energy spectra reconstructed with neural networks compared to the results of PAMELA (a) and AMS-02 (b) measurements. Panels (c) and (d): same for helium energy spectra.}
\end{figure*}

The full trained models reconstruct GCR flux time series at different energies for 2019--2022. While daily AMS-02 data are unavailable for this interval, BR-averaged values have been published~\citep{Aguilar_2025, Aguilar_2025-1}. The truncated models reconstruct GCR spectra for 2006--2011, for which 1--3-month-averaged proton~\citep{Adriani_2013, Martucci_2018} and helium~\citep{Marcelli_2020, Marcelli_2022} fluxes from PAMELA have been published. Figure~\ref{fig_all_time_series} presents the particle flux time series reconstructed by the neural networks and by the NM YF+FFA method, along with the experimental data.

For 2019--2022, good agreement is observed between AMS-02 data and neural-network predictions. The sole exception is high-energy helium fluxes (Figure~\ref{fig_AMS_he_high}), for which the reconstructed values systematically exceed the measurements. However, this discrepancy likely arises because daily and BR-averaged AMS-02 data use different energy binning (see Figure~\ref{fig_spectra}), so the compared fluxes do not correspond exactly in energy. For the FFA-based reconstructions, the same issues observed in 2011--2019 persist: at low energies (Figures~\ref{fig_AMS_p_low} and~\ref{fig_AMS_he_low}), this approach overestimates GCR variation amplitudes, while at high energies (Figures~\ref{fig_AMS_p_high} and~\ref{fig_AMS_he_high}), it predicts nearly constant fluxes.

For 2006--2011, good agreement is observed between the neural-network-reconstructed low-energy helium flux and PAMELA data (Figure~\ref{fig_PAMELA_he_low}); however, significant differences appear in other cases. The reconstructed low-energy proton flux in 2009--2011 (Figure~\ref{fig_PAMELA_p_low}) lies approximately 10\% below PAMELA measurements. However, the PAMELA collaboration noted that at the lowest energies, their proton fluxes were about 10\% higher than those from AMS-02 during the overlapping period~\citep{Martucci_2018}. Since our models were trained on AMS-02 data, this fact explains the discrepancy. Differences at the 10\% level at high energies (Figures~\ref{fig_PAMELA_p_high} and~\ref{fig_PAMELA_he_high}) were also observed during initial PAMELA--AMS-02 comparisons~\citep{Marcelli_2022}. Thus, accounting for instrumental differences, the agreement between reconstructed and experimental fluxes can be considered satisfactory. The model effectively reproduces short-term variations; for instance, the recurrent 27-day variations at the 2007--2008 transition, studied using unpublished PAMELA data~\citep{Krainev_2018, Modzelewska_2020}, are traceable to the highest energies.

Figure~\ref{fig_spectra} presents the results as energy spectra at fixed times rather than as fixed-energy time series. Instead of force-field results, the experimental data are compared with predictions from both CNN and MLP models. The neural networks show nearly perfect agreement on these datasets. However, proton fluxes reconstructed for 2009 lie below PAMELA data up to approximately 4~GeV/nuc (Figure~\ref{fig_PAMELA_p_spectrum}).

Besides PAMELA--AMS-02 differences, another reason may be that the sunspot numbers and NM count rates observed during the solar cycle 23/24 minimum fall outside the training dataset range. Consequently, the neural networks must extrapolate, which can lead to inaccuracies. For this reason, random-forest and gradient-boosted-tree models were discarded, as decision trees fundamentally cannot extrapolate beyond their training range.

\subsection{Correlation between the time series reconstructed with two methods\label{correlation}}

With time series of particle fluxes reconstructed by two different methods, we can analyze correlations between their variations. Given that the machine-learning-based predictions agree well with direct measurements, this analysis effectively reveals the inaccuracies of the FFA-based approach.

\begin{figure}
\centering
\begin{minipage}[b]{0.508\linewidth}
    \begin{subfigure}[b]{\textwidth}
\includegraphics[width=\linewidth]{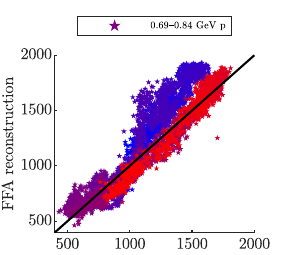}
  \caption{\label{fig_CorrelationsFull_a}}
\end{subfigure}
\begin{subfigure}[b]{\textwidth}
\includegraphics[width=\linewidth]{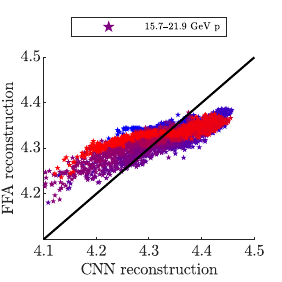}
  \caption{\label{fig_CorrelationsFull_b}}
\end{subfigure}
\end{minipage}
\begin{minipage}[b]{0.462\linewidth}
    \begin{subfigure}[b]{\textwidth}
\includegraphics[width=\linewidth]{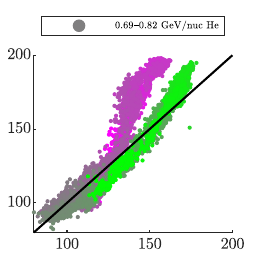}
  \caption{\label{fig_CorrelationsFull_c}}
\end{subfigure}
\begin{subfigure}[b]{\textwidth}
\includegraphics[width=\linewidth]{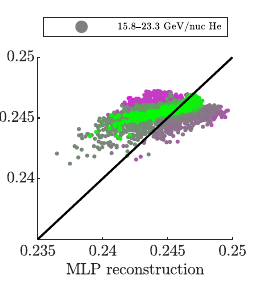}
  \caption{\label{fig_CorrelationsFull_d}}
\end{subfigure}
\end{minipage}
\caption{\label{fig_correlations} Scatter plots showing the relationship between GCR fluxes reconstructed with the force-field approximation and with neural networks. Panel (a): low-energy proton flux; panel (b): high-energy proton flux; panels (c) and (d):  low-energy and high-energy helium fluxes, respectively.}
\end{figure}

Figure~\ref{fig_correlations} shows scatter plots of fixed-energy fluxes reconstructed for the entire 2006--2022 period. The color gradient encodes time. For protons (Figures~\ref{fig_CorrelationsFull_a} and~\ref{fig_CorrelationsFull_b}), points from 2006 are bright blue; as solar activity decreases toward the 23/24 minimum (2009), the flux reaches its maximum (dark blue). During the subsequent rise, the color shifts to purple; between the cycle-24 maximum (2014) and the 24/25 minimum (2020), it evolves from purple to dark red; during the rise in 2020--2022, it becomes bright red. For helium (Figures~\ref{fig_CorrelationsFull_c} and~\ref{fig_CorrelationsFull_d}), the palette evolves analogously (from magenta to green).

At low energies, a pronounced hysteresis appears between the results of two reconstructions: a given flux predicted by the neural network corresponds to two branches of FFA-based values~--- one before 2014 and the other after. This pattern resembles the hysteresis reported in~\citep{Aguilar_2025}, where a similar effect between opposite charge signs was attributed to the 2013--2014 heliospheric polarity reversal.

For reference, the line $y = x$, which corresponds to exact agreement at all solar activity levels, is also shown in all panels of Figure~\ref{fig_correlations}. For low-energy protons, the post-2014 branch ($A>0$) lies relatively close to this line, deviating mainly at the highest and lowest activity levels (FFA exhibits stronger variations). The pre-2014 branch ($A<0$) lies entirely above the line, indicating that the force-field approximation systematically overestimates the flux relative to the neural-network reconstruction. The same features are present for low-energy helium, and are even more pronounced. In case of protons, this phenomenon was demonstrated in~\citep{Vais_2025}, but the authors of this paper compared FFA-based reconstruction results to AMS-02 observations directly, so the pre-2014 branch had less than 3 years of data and therefore was short. The present study supports these findings and extend them to the case of prolonged time series and modified FFA.

At high energies, hysteresis persists but is less distinct due to weaker modulation and larger random fluctuations. For both protons and helium, the clouds are more horizontally elongated, indicating that fluxes reconstructed with NM YF+FFA approach fluctuate weakly around their mean at these energies.

Correlation coefficients between the time series reach $\approx 0.92$ for the lowest-energy protons, then increase to a peak of $\approx 0.96$ for rigidities $R \in [4, 11]$~GV, and subsequently drop rapidly, so it is equal to 0.67 for bin number 27 ($R=22.8-33.5$~GV) and reaches 0.36 for the last bin ($R=69.7-100$~GV). For helium, the rigidity dependence is similar but coefficients are lower by 2--4\%. If only the right hysteresis branch (post-2014) is considered, the correlation does not fall below 0.97 up to $R = 15$~GV and peaks at 0.99 for both species. Thus, the modification of force-field approximation used here describes the solar modulation of proton and helium spectra well during periods of positive heliospheric polarity, whereas the properties of modulation under negative polarity need further investigation.

\section{Conclusion \label{Conclusion}}

In this work we reconstructed time-resolved GCR proton and helium spectra using two complementary approaches: a neutron monitor yield-function method coupled with a force-field parameterization, and data-driven models based on artificial neural networks.

The yield-function-based framework was calibrated to AMS-02 data, including a refined accounting for heavy-nuclei contributions and the helium isotopic composition, and tested across solar-cycle phases. For the first time, we applied a two-parameter version of the force-field approximation to neutron monitor data, as well as tried to calibrate neutron monitor yield-function using an additional parameter. While this physics-guided approach recovered long-term trends, it systematically overestimated short-term variability at low energies and underestimated modulation at high energies, and exhibited polarity-dependent hysteresis relative to data-driven reconstructions.

Neural networks trained on multi-station neutron monitor count rates, heliospheric magnetic field polarity, sunspot number, and Ap index (with physically motivated delays) achieved markedly better fidelity: they reduced mean absolute percentage error by factors of about three and maintained $\chi^2/\mathrm{dof}$ typically between 0.1 and 2. Also, it appears that neural networks seamlessly reproduce effects of short-term solar modulation, but this topic requires further investigation. SHAP analysis confirmed physically consistent station–energy contributions linked to geomagnetic cutoff rigidity and limited relevance of Ap index outside major magnetic storms. Applying full and truncated models, we extended spectra to 2019--2022 (in agreement with AMS-02 BR-averaged data) and to 2006--2011 (showing overall consistency with PAMELA within known inter-experiment systematics and recognizing extrapolation limits during the deep 23/24 minimum).

Together, these results demonstrate that the global neutron monitor network can be utilized as a unified GCR spectrometer and highlight where force-field parameterization requires refinement, particularly in low-energy range and during negative-polarity epochs.

\section*{Acknowledgments}
We gratefully acknowledge the PIs of the NM stations whose data was used in this study, namely BRBG, MRNY, SOPO, THUL, APTY, TXBY, OULU, KERG, YKTK, MOSK, NVBK, LMKS, JUNG, AATB, MXCO, ATHN and PSNM. We acknowledge the NMDB database, founded under the European Union's FP7 programme (contract no. 213007), and IZMIRAN database for providing NMs data.  Daily values of sunspot number and Ap index are provided by WDC-SILSO, Royal Observatory of Belgium, Brussels and Geomagnetic Observatory Niemegk, GFZ German Research Centre for Geosciences, Potsdam, respectively. Solar polar field data are provided by Wilcox Solar Observatory. Mean atmospheric depths at the NM locations are calculated with MATLAB Aerospace Toolbox. The authors utilized large language models (in particular, ChatGPT and DeepSeek) in order to check grammar and punctuation during writing.

\section*{Author Contributions} 

S. Siruk administrated the project, prepared, visualized and synthesized data, wrote the initial version of the paper. 
V. Alekseev and V. Kuzminov developed software and performed formal analysis of the data.
R. Yulbarisov validated the results, reviewed and edited the manuscript. 
A. Mayorov conceptualized the research, acquired the financial support and supervised the project.

\section*{Funding}
This work was supported by the Russian Science Foundation under Project ``Study of the charge dependence of solar modulation of galactic cosmic rays'' (project no. 25-22-00508). 

\section*{Conflicts of Interest}

The authors declare that there is no conflict of interest regarding the publication of this article.

\section*{Data Availability}
All neural-network-based models developed in this study, along with an application service, are available in a public repository~\footnote{\url{https://github.com/VictorAboba/Spectrum_nn_app.git}}.

\bibliographystyle{aasjournal}
\bibliography{sample}

\end{document}